\documentclass[12pt,letterpaper]{article}
\usepackage{amsmath,pgf,pgfarrows,pgfnodes,float,appendix, hyperref,scalerel,amssymb}
\usepackage{graphicx}
\usepackage{subfigure}

\usepackage[margin=0.9in]{geometry}
\usepackage{pgfplots}
\usepackage{amsmath}
\usepackage{tikz}

\title{{\rm\footnotesize \qquad \qquad \qquad \qquad \qquad \ \qquad \qquad \qquad \ \ \ \ \ \                   }\vskip.5in  Fluctuations and Correlations in Causal Diamonds}
\author{Tom Banks\\ 
Department of Physics and NHETC\\
Rutgers University, Piscataway, NJ 08854\\
E-mail: \href{mailto:tibanks@ucsc.edu}{tibanks@ucsc.edu}
\\
\\
Willy Fischler\\
Department of Physics and Texas Cosmology Center\\
University of Texas, Austin, TX 78712\\
E-mail: \href{mailto:fischler@physics.utexas.edu}{fischler@physics.utexas.edu}}
\date{}
\begin{document}
\maketitle

\begin{abstract} 
We argue that the correlation between the modular Hamiltonians of two causal diamonds related by a finite time translation is equal to the uncertainty in the modular Hamiltonian of the overlap diamond in empty Minkowski space-time. 
In a separate section we investigate the derivation of length fluctuations in Minkowski diamonds from the 't Hooft commutation relations, and the connection of those fluctuations to entropy fluctuations.  
\end{abstract}

\section{Introduction}

In\cite{VZ1}, Verlinde and Zurek made the fascinating suggestion that quantum gravitational fluctuations might be measurable in modern interferometer experiments.  A variety of subsequent papers\cite{subsequent} have demonstrated that large fluctuations of the sort suggested in\cite{VZ1} indeed occur in AdS/CFT, or follow from other plausible theoretical conjectures.  At present we do not have a calculation, based on some sort of fundamental principles, of the actual observables relevant to interferometer experiments.  The aim of the present paper is to make incremental progress towards that goal.  The basic interferometer observable is the time correlation function of arrival time fluctuations in single passes of the interferometer separated by a proper time interval $\tau$ at the beam splitter.   We will instead calculate the time correlation function of the modular Hamiltonians of two causal diamonds along the same geodesic in Minkowski space, with proper time separation $\tau$ .  We will argue that when there is no overlap between the diamonds, the correlation function is essentially zero, and that it is otherwise equal to the fluctuation of the modular Hamiltonian of the overlap diamond .  

In the second part of the paper we will try to relate the modular fluctuation to the 't Hooft commutation relations\cite{tHcr} which are also related to the fluctuation of lengths along individual interferometer trajectories studied in\cite{VZ1}.   Our formulae differ from those found in\cite{VZ3}.   We are unable to calculate precise numerical coefficients for either the modular or the length fluctuations using these methods.   In particular, we cannot check the relation\cite{Carlip}\cite{solo}\cite{BZ}
\begin{equation}   (\Delta K)^2 = \langle K \rangle . \end{equation}  However we argue that if the Planck scale regularization of the transverse geometry of causal diamonds is "fuzzy", then the calculation of entropy fluctuations from length fluctuations is compatible with this formula and the normalization of length fluctuations "at the same angle" is fixed by this entropy fluctuation coefficient.  This leads us to a conjecture about the time correlation function of length fluctuations, which might have direct relevance to interferometer experiments.  
 
 \section{Time Correlations in terms of Equal Time Fluctuations}
 
 The basic observable in an interferometer experiment is the correlation between arrival times of signals emitted from the beam splitter in different passes of the apparatus, separated by proper time $\tau$ at the beam splitter.  In the interpretation of these signals in terms of changes in the space-time metric, these are interpreted as distortions of the length of the arms of the interferometer.  We are interested in possible quantum gravitational fluctuations of these lengths in an empty causal diamond in Minkowski space.  As usual, we neglect the back-reaction of the measuring apparatus on the causal diamond in which it resides.   A related, and more theoretically tractable quantity to calculate is the time dependent correlation of modular Hamiltonians for two causal diamonds along the same Minkowski geodesic, separated by proper time $\tau$.  A pictorial description of the setup is shown in Fig.1 .   There are two causal diamonds, $D_{1,2}$ which may or may not overlap,  and a large diamond $D_L$ which exactly contains them both.  The modular Hamiltonians for those three diamonds are denoted $K_1, K_2, K_L$ respectively.   The quantity we wish to calculate is
 \begin{equation} G(\tau) \equiv {\rm Tr}\ [e^{-K_L} (K_1 K_2)] -  {\rm Tr}\ [e^{-K_L} K_1] {\rm Tr}\ [e^{-K_L} K_2],\end{equation} as a function of the proper time difference $\tau$ between the past tips of $D_1$ and $D_2$.  
 
  \begin{figure}[h]
\begin{center}
\includegraphics[width=01\linewidth]{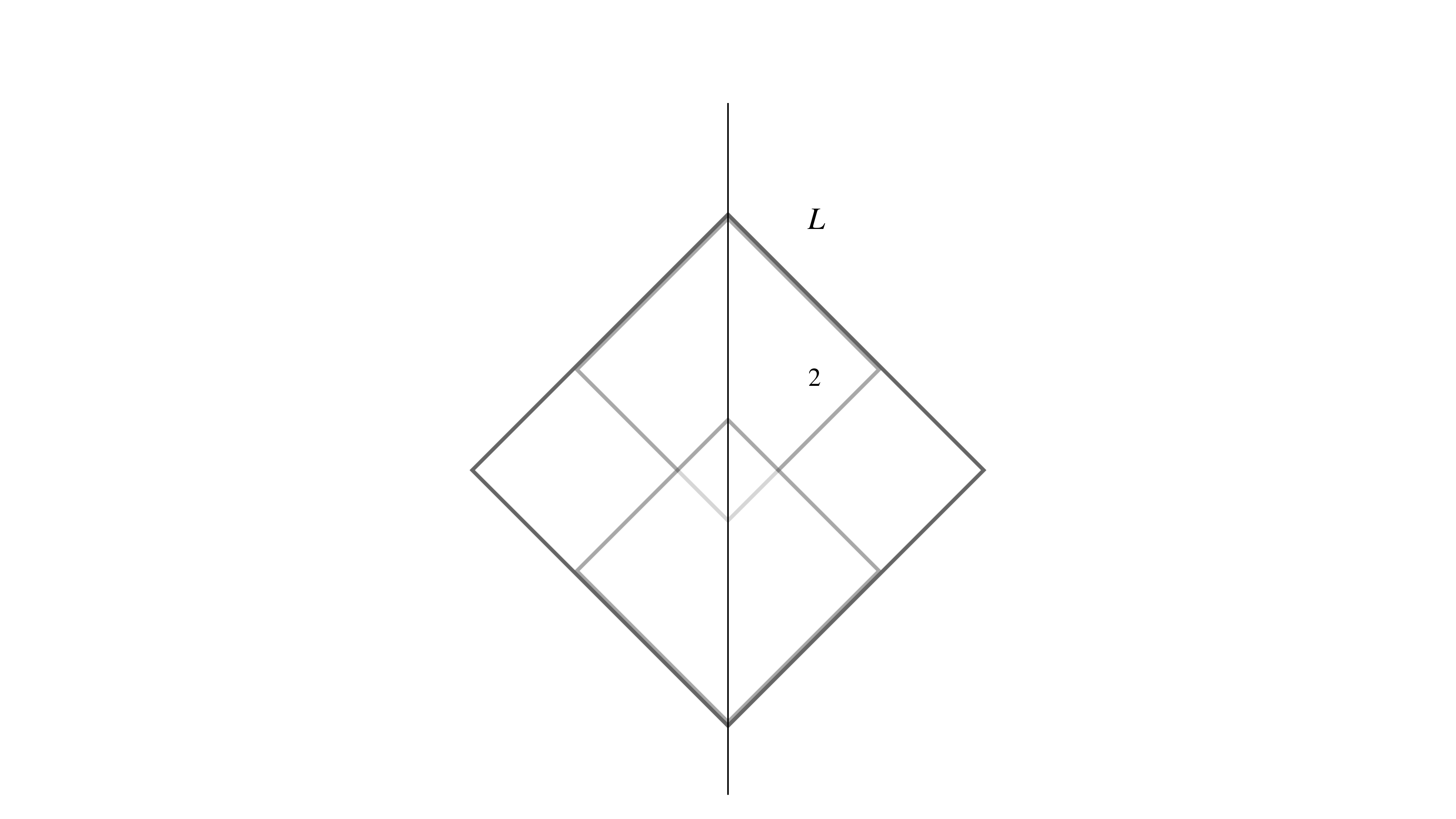}

\caption{Shifted Diamond Correlation Function.} 
\label{fig:shifteddiamond}
\end{center}
\end{figure}

 We will assume some basic principles in order to perform this calculation.  These principles have been described in\cite{hilbertbundles}, but in fact we will need much less than the full set of assumptions of that paper in order to establish our result.   Any causal diamond can be covered by both a past and future directed nested set of diamonds Fig.2.
  \begin{figure}[h]
\begin{center}
\includegraphics[width=01\linewidth]{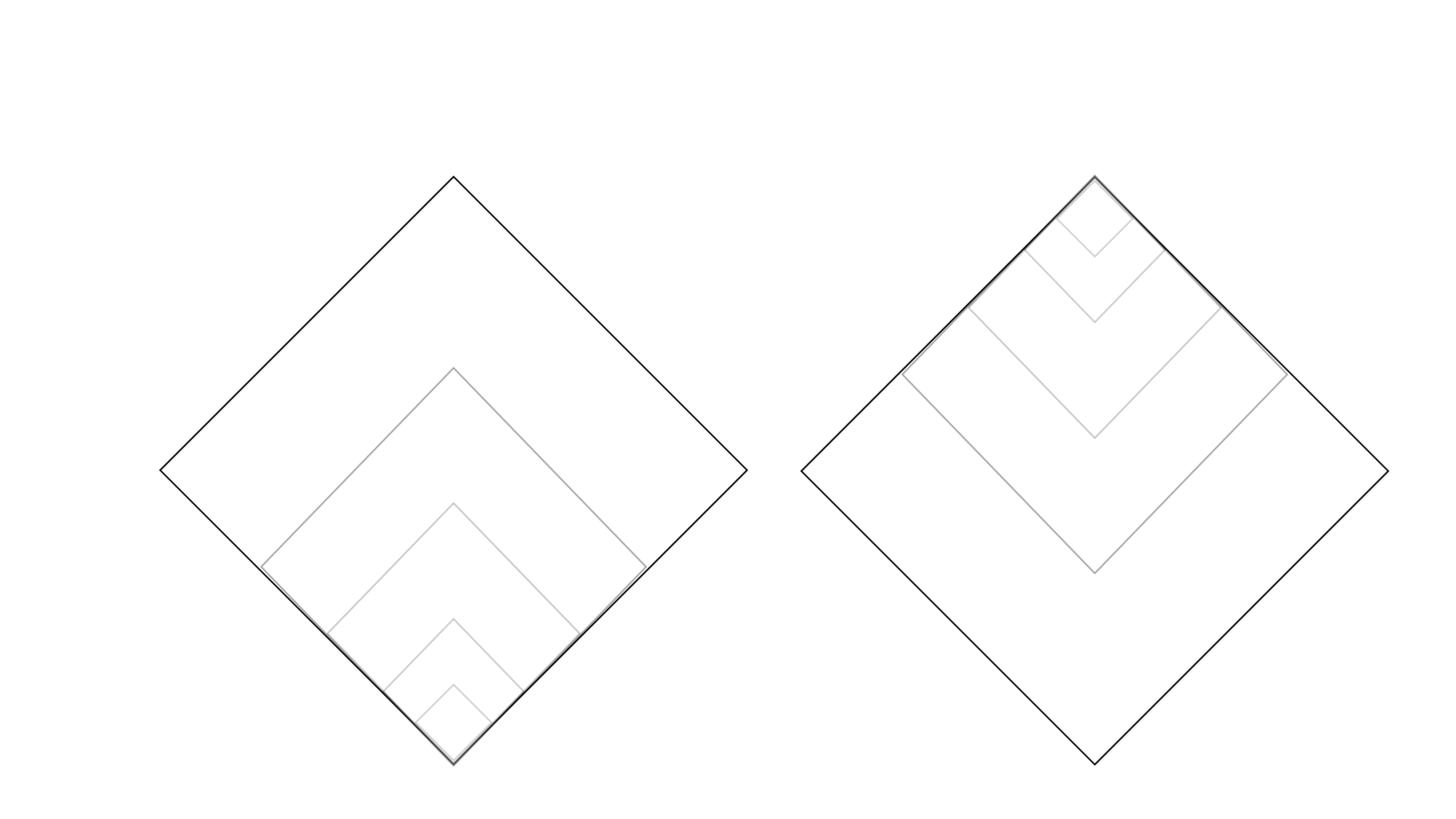}

\caption{Future and past directed nested coverings of a causal diamond.} 
\label{fig:nesteddiamondsforcorrelation}
\end{center}
\end{figure}

   The time differences between the future (past) tips of the future (past) directed nesting are all $L_P$.  If we consider the operator $e^{iK(t)} e^{- iK(t \pm L_P)}$, for each pair of sub-diamonds in a future/past directed nesting, this defines a sequence of unitary embeddings of the Hilbert space of the smaller diamond into the larger diamond.  We can view this as a {\it choice} of time slicing in Lorentzian signature Minkowski space.  Here, we're implicitly assuming all Hilbert spaces are finite dimensional and that the log of their dimension is roughly given by the Bekenstein-Hawking-Jacobson-Bousso area law.  This implies in particular that a single Planck step in time gives rise to an exponentially large change in dimension, once the dimension is macroscopic.  
 
 The other important assumption we will make is the {\it Milne red-shift}: we assume all of the quantum degrees of freedom in an empty causal diamond reside on the {\it stretched horizon} of the diamond.   The time evolution we are describing is proper time along the geodesic traversed by the beam splitter.  Dynamics on the stretched horizon of a diamond of size $L$, has a time scale of order $L$ in geodesic proper time.  
 
 We can now proceed to calculate $G(\tau)$.  We first consider the case where $D_1$ and $D_2$ have no overlap.  In this case, the Hilbert space of $D_L$ is exponentially larger than either of the subsystem Hilbert spaces.  Page's theorem\cite{page} says that, for a Haar random state on the Hilbert space of $D_L$ the states on the subsystem Hilbert spaces are exponentially close to maximally uncertain, which implies vanishing of $G(\tau)$: there is no correlation between the two subsystems.  The modular Hamiltonian on the Hilbert space of $D_L$ is, according to\cite{Carlip}\cite{solo}\cite{BZ}\cite{hilbertbundles},  the $L_0$ generator of an interacting two dimensional CFT, with interactions that mix up the individual degrees of freedom describing the subsystems $D_i$ in a complicated way.  While we have no simple proof of the analog of Page's theorem in this case it seems extremely likely that $G(\tau)$ will vanish in this case as well.
 
 Another limiting case is $\tau = 0$.   In this case $D_L = D_1 = D_2$ and $G(0) = (\Delta K_1)^2$ the uncertainty, or fluctuation in the modular Hamiltonian of either of the two subsystems.   These two limits motivate the following general conjecture:  
 \begin{equation} G(\tau) = {\rm Tr}\ [e^{- K_{12} (\tau)}  (K_{12} (\tau)  - \langle K_{12} (\tau) \rangle)^2] , \end{equation} where $K_{12} (\tau)$ is the modular Hamiltonian of the overlap diamond, and the angle brackets mean expectation value in the density matrix $e^{-K_{12} (\tau)}$.  We will "prove" this conjecture by a sort of "continuous mathematical induction".  We assume that it is true for some value of $\tau$ and then argue that it continues to hold for $\tau + \delta \tau$ for small enough $\delta \tau$.   Since it holds for $\tau = 0$ and for $\tau > 2T$, the total proper time in the diamonds $D_{1,2}$, this establishes the conjecture.
 
 Our argument is pictorially described in Fig.3 .  
  \begin{figure}[h]
\begin{center}
\includegraphics[width=01\linewidth]{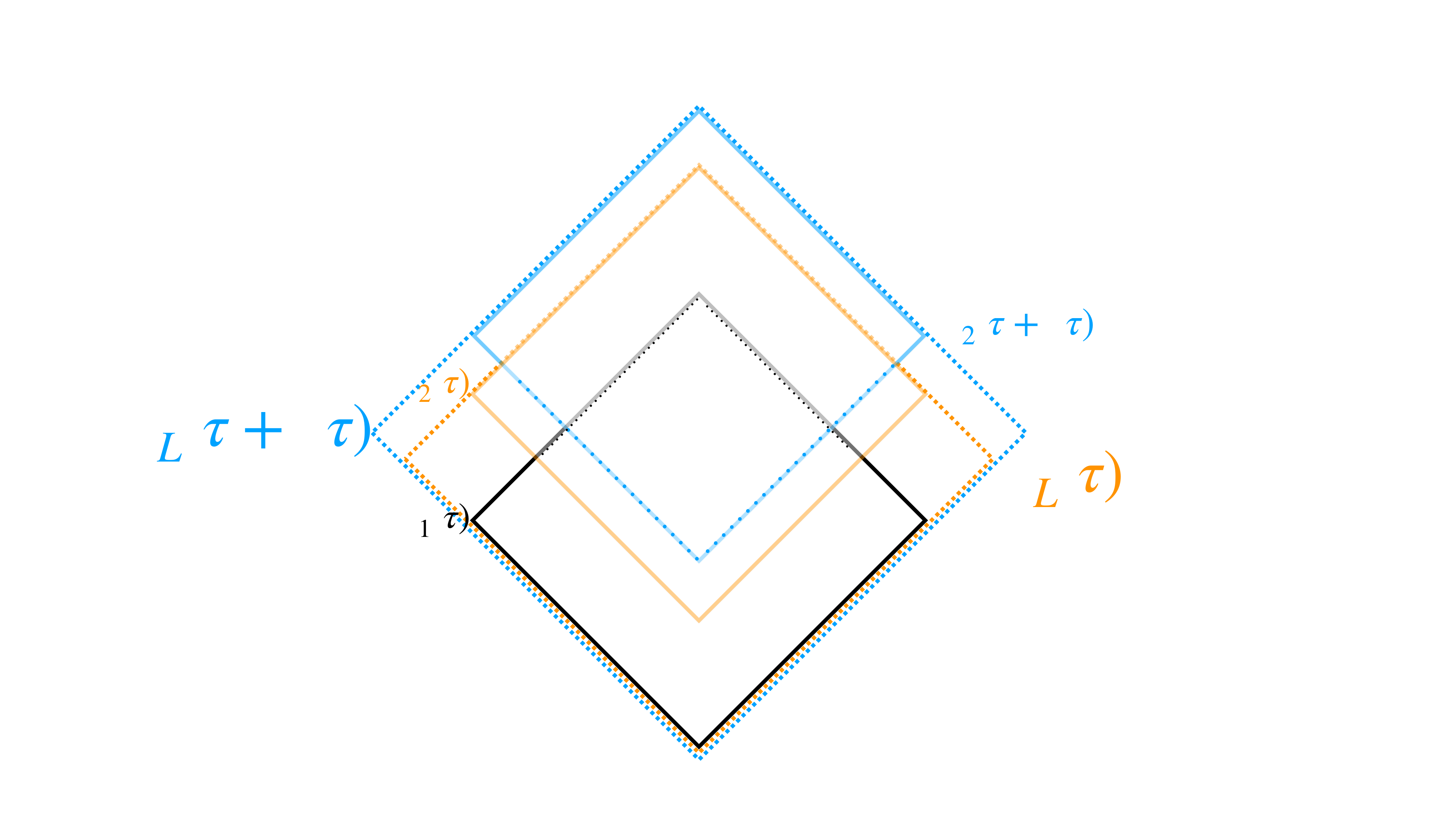}

\caption{Infinitesimal Diamond Shift.} 
\label{fig:infinitesimaldiamondshift}
\end{center}
\end{figure}

 We are comparing the situation where the diamond $D_2$ is the orange diamond in this figure, to that in which it is the blue diamond, shifted forward in time by $\delta \tau$.  Two things have occurred.   The diamond $D_L$ has grown, by the addition of degrees of freedom that are associated with the region between the blue and orange dotted lines.   The overlap diamond $D_{12} (\tau + \delta \tau) $ has shrunk, losing degrees of freedom associated with the region labeled $\delta$ in the figure.   Our assumption about the relation between modular Hamiltonians, time evolution and embedding maps, combined with the Milne red shift, implies that the new degrees freedom in $D_L$ cannot yet be correlated with those in $D_1$ and do not contribute to $G(\tau)$.   The degrees of freedom in region $\delta$ {\it are} causally connected to the new degrees of freedom in $D_L$, though not to those in the shifted (blue) $D_2$.  Since they travel with the speed of light nothing prevents them from being entangled with those new variables.   Monogamy of entanglement then implies that they are no longer entangled with the variables in $D_1$ and so cannot contribute to $G(\tau)$.  
 
 By assumption, $G(\tau)$ before the shift by $\delta \tau$ was just the modular fluctuation of $K_{12} (\tau)$.  We have just argued that the entire effect of the shift is simply to disentangle a small fraction of the total number of variables in the Hilbert space of $D_{12} (\tau)$ leaving only those variables associated with the smaller overlap.  In the explicit models of\cite{hilbertbundles} we are just removing a small fraction of the total number of fermion fields in a Thirring model involving the product of two commuting $U(1)$ currents.   $G(\tau)$ will still be given by the modular fluctuation in the smaller model.
 
 \section{Modular Fluctuations and 't Hooft Commutators}
 
 In 1998, Carlip and Solodukhin argued that the effective theory on black hole horizons, for non-extremal black holes in space-times with non-negative cosmological constant (c.c.) was a (cut-off) $1 + 1$ dimensional CFT on an interval.  In\cite{BZ} this argument was generalized to a large class of causal diamond boundaries.  It is even a correct description of boundary anchored Ryu-Takayanagi diamonds in AdS space, or of single nodes in a tensor network description of large AdS black holes in CFTs with a large radius AdS dual.   For any such diamond boundary, the modular fluctuations satisfy
 \begin{equation} (\delta K)^2 = \langle K \rangle . \end{equation}
 Recently, Verlinde and Zurek\cite{VZ3} have presented a derivation of this result based on 't Hooft's shock wave commutation relations.  This is extremely interesting because it points to a possible connection between the more easily accessible entropy fluctuation calculations and the angular dependent correlations necessary for making connection to actual experiments.   The present section is devoted to our attempts to forge similar connections.
 
 We will first present a calculation of the equal time two point functions
 \begin{equation} \langle X^{\pm} (t , \Omega)  X^{\pm}( t), \Omega^{\prime}) \rangle , \end{equation}  where we will explain below the meaning of the angle brackets.    We then use these in two ways.  First, following\cite{VZ1} we calculate the length fluctuations along a single light ray trajectory in a causal diamond.  Then, combining such calculations for many angles, we calculate the entropy fluctuations.  Both calculations give finite answers but the entropy calculation has an ambiguous normalizations having to do with how we discretize things at the Planck scale.   The normalization of the entropy calculation can be fixed by invoking the C-S answer.   
 
 We begin from 't Hooft's commutation relations for the fluctuating null coordinates near the boundary of a causal diamond in Minkowski space\cite{tHcr}\cite{V293}.  Near the diamond boundary the largest components of Einstein's equations are those in the two nearly null directions.  They imply that the transverse metric $G_{ij}$ is independent of those null coordinates and that the two dimensional Lorentzian metric
 \begin{equation} \hat{g}_{ab} = g_{ab} - g_{ai}G^{ij} g_{bj} , \end{equation} is flat.   The fluctuations away from flat metrics are parametrized by a conformal factor and a coordinate transformation from the Minkowski form $\eta_{ab} L_P^2$.  Setting the conformal factor to zero we get
 \begin{equation} \hat{g}_{ab} = \eta_{ab} L_P^2 + L_P (\eta_{\alpha b} \partial_{\beta} X^b + \eta_{a \beta} \partial_{\alpha} X^a) \equiv \eta_{ab} + h_{ab} . \end{equation}  
 As shown in\cite{V293}, when this is plugged into the Einstein action, we get a topological action which is a pure boundary term and leads to the commutation relations
 \begin{equation} [X^+ (t, \Omega), X^- (t , \Omega^{\prime})] = i  (J^2 + 1)^{-1} (\Omega, \Omega^{\prime}) L_P^2 . \end{equation}
 Here $J^2 $ is the squared angular momentum operator on the two sphere.
 Note that we are using conventions where metrics have dimensions and coordinates are dimensionless.  Also, the parameter $t$ is $x^+$ on the past boundary and $x^-$ on the future boundary.  Furthermore, there is always a gauge choice in which $X^+$ vanishes on the past boundary, and $X^-$ on the future.    Thus, this commutation relation applies only on the holographic screen or bifurcation surface of the diamond boundary.   Note however that every point on the diamond boundary lies on the bifurcation surface of {\it some} diamond in either a past or future directed nested covering of the diamond by smaller diamonds.   Thus, we have an 't Hooft commutation relation at every point on the diamond boundary, but they refer to operators that operate in different Hilbert spaces.   We should also emphasize that these commutation relations are {\it incompatible} with our fundamental assumption that the diamond Hilbert spaces are finite dimensional.  We view them as approximate statements, which need regularization.  In\cite{hilbertbundles}, one of us proposed a regularization of these operators in terms of approximate $U(1)$ currents in cut-off CFTs.  In the present paper we will work with the infinite dimensional algebras, but use them only to calculate the Gaussian fluctuations in the $X^{\pm}$ variables.
 
 In order to calculate correlation functions, we must decide on the state representing an empty diamond.  Fortunately, symmetry considerations are enough to determine the state.   Every empty diamond in Minkowski space is time reversal invariant.   Time reversal takes the canonically conjugate variables $X^{\pm}$ into each other.  This implies that the uncertainties of the angular momentum components of $X^{\pm}$ must be identical, but that implies that the state is the minimal uncertainty state for both $X^+$ and $X^-$.  
 \begin{equation} (\Delta X^{\pm}_{kl})^2  = \frac{(l(l+1) + 1)^{-1}}{2} L_P^2 . \end{equation} 
 Furthermore, the mixed two point function between $X^+$ and $X^-$, vanishes.     We conclude that
 \begin{equation} \langle X^{\pm} (t , \Omega) X^{\pm} (t, \Omega^{\prime}) \rangle = \frac{1}{2} (J^2 + 1)^{-1} (\Omega, \Omega^{\prime}) L_P^2 . \end{equation}
 It is convenient to work with the angular momentum decomposition of these relations, because this allows us to incorporate the finite entropy assumption that we have always used in the Holographic Space Time (HST) formalism.  The variables in HST are sections of the spinor bundle on the two sphere, with an angular momentum cutoff.  In the updated version of HST compatible with the Carlip-Solodukhin ansatz\cite{hilbertbundles}, the expansion coefficient multiplying each angular momentum component of the transverse spinor is a one plus one dimensional massless fermion field, living on an interval (the stretched horizon) near the boundary of the causal diamond, with a UV cutoff.   The UV cutoff is chosen so that the entropy of the smallest transverse angular momentum field theory coincides with that of the smallest causal diamond to which the semi-classical CS analysis applies.  According to our present understanding, only some far future experiment might tell us what that size is.  However, the use of Cardy's theorem by CS allows us to put a plausible lower bound on it.  Cardy's theorem gives a more or less correct count of the number of states of a single Dirac fermion on an interval when it has between $10-15$ momentum modes.  
 
 Larger causal diamonds are modeled by adding angular momentum multiplets, according to the CS rule.  We hypothesize that the $X^{\pm}$ variables are bilinears in these fermion fields, and so carry a maximal angular momentum equal to twice that of the spinors.   Both these maximum values scale like the proper time $2T$ in the diamond.
 
 The length fluctuation along a light ray trajectory on the diamond boundary has to be treated with great care.  Formally, the length is the sum of integrals of $h_{++}$ along the past portion of the trajectory and $h_{--}$ along the future portion, and the metric variables are derivatives of the $X^{\mp}$ variables.  However, we have just argued that the $X^{\pm}$ variables at different spheres on the boundary, which are the holographic screens of different diamonds in a nested cover, are {\it independent random variables}.  Thus $X^{\mp}$ are nowhere differentiable functions of their time arguments.
 Instead of simply integrating a total derivative, we divide proper time along the geodesic into steps of size $L_P$ and define the infinitesimal length fluctuation on the corresponding segment of the boundary trajectory to be the difference of the $X^{\pm} $ variables at those two points along the trajectory.  The length fluctuation thus has an angular momentum decomposition
 \begin{equation} \langle L_{Jm} (i) L_{Jm} (i) \rangle = (J^2 + 1)^{-1} L_P^2  , \end{equation} twice as large as that of the individual 't Hooft variables.  Furthermore, the length fluctuations for independent segments of the trajectory are independent random variables, because they each contain an $X^{\pm}$ variable not shared by any previous segment.   Thus, the total length, which is the sum of these individual randomly fluctuating interval lengths, satisfies
  \begin{equation} \langle L^{T}_{Jm} (i) L_{Jm} (i) \rangle = \frac{2T}{L_P} (J^2 + 1)^{-1} L_P^2  .\end{equation}
  Note that there is an additional hidden $T$ dependence in this formula, because the maximal angular momentum grows like $T$.  If one were really contemplating light rays that were delta function localized in angle, this would lead to an extra logarithmic enhancement of the length fluctuation.  As we'll see below, the constraint that the entropy of a finite area diamond is finite leads to the conclusion that its holographic screen has a pixel of minimal angular area .  The length fluctuation of a light beam should be averaged over this pixel in some way.  Consistency with the Carlip-Solodukhin formula for entropy fluctuations will imply that the logarithmic enhancement of the naive formula has a cutoff at a fixed finite value, related to the UV cutoff on the $1 +  1$ dimensional fermion fields.
 We'll see that it is natural to obtain a finite value because "pixels" made of a finite number of angular momentum modes have only "fuzzy" localization, but that there is no obvious "correct" way to define the pixelization.

  \subsection{Entropy fluctuations from the 't Hooft Commutation relations}
  
  Carlip\cite{Carlip} and Solodukhin\cite{solo} (C-S) showed that the fluctuations around the horizon of generic black holes contained the hydrodynamic equations of a $1 + 1$ dimensional conformal field theory (CFT) living on a strip (the stretched horizon), with central charge proportional to the horizon area.  Using Cardy's formula they derived the Bekenstein-Hawking area formula. The authors of\cite{BZ} pointed out that the C-S derivation worked for arbitrary causal diamonds\footnote{To understand large AdS black holes one needs to invoke the tensor network description of AdS space.  For CFTs with an Einstein-Hilbert dual, he C-S argument works at the level of a single node of the tensor network, but the entropy of large black holes is dominated by excitations that couple many nodes together. The fluctuation formula for the modular Hamiltonian of large AdS black holes is that of a higher dimensional CFT.} and implied the universal fluctuation formula for the modular Hamiltonian
  \begin{equation} \langle (K - \langle K \rangle)^2\rangle = \langle K \rangle . \end{equation} This had been shown\cite{VZ2}\cite{perl} to hold for boundary anchored Ryu-Takayanagi diamonds in arbitrary AdS/CFT models with Einstein-Hilbert duals.  
  
  The description of the geometry of the holographic screen of a diamond in terms of a $1 + 1$ CFT arose independently in our model of holographic cosmology\cite{holocosm}.  There, the CFT was naturally realized in terms of $1 + 1$ dimensional fermion fields, which are the expansion coefficients of a spinor field on the holoscreen in terms of eigenspinors of the Dirac operator.  The idea of describing Riemannian geometry in terms of the Dirac operator is due to Connes\cite{connes}.
  Finite central charge implies a UV cutoff on the Dirac eigenvalue.  Finite {\it entropy} implies that one should also cut off the number of momentum modes of the individual fermion fields on the stretched horizon.  Mathematically it's natural to take this cutoff to be at the point where Cardy's formula begins to be accurate, which is about $10-15$ momentum modes for a single fermion field.  Physically this corresponds to the smallest diamond in Planck units for which one can believe the semi-classical, hydrodynamic arguments of C-S.  Ultimately the correct value of this cutoff in a theory applying to the real world would have to be taken from experiment.   
  
  In Minkowski space, the cutoff on the transverse Dirac eigenvalue is a cutoff on angular momentum.  The expectation value of the radius of the holographic screen is proportional to this cutoff, implying a UV/IR connection like Maldacena's scale/radius duality.  Fluctuations in the unimodular geometry of the screen are described by fluctuations of the two dimensional fermion fields, which are the expansion coefficients of the spinor field on the holographic screen in terms of Dirac eigenspinors on the classical screen geometry.  
  
  In this paper however, we want to make a more semi-classical estimate of fluctuations, using the 't Hooft commutation relations.  The cutoff on angular momentum implies that we have only $\sum_{1/2}^{L_M} (2 L + 1)$ angular momentum modes of the spinor fields on the holographic screen.  Using the Bekenstein-Hawking-Jacobson-Bousso formula we have 
  \begin{equation} \frac{A}{4G} \approx 2K \sum_{1/2}^{L_M} (2 L + 1) {\rm ln}\ 2 . \end{equation} where $K$ is the number of momentum modes of each $1 + 1$ dimensional Dirac field on the stretched horizon.   
  In\cite{hilbertbundles} it was suggested that the 't Hooft operators are constructed out of $U(1)$ currents bilinear in the underlying fermions.  These have angular momenta that range from $0$ to $2L_M$.   Thus the number of "pixels" into which we can break up our screen is $\sum_{L = 0}^{2L_M} (2L + 1)$.   A pixel, by definition, is the best approximation of the delta function on the smallest angular subset of the sphere that we can make out of those spherical harmonics.  There are of course many ways to define "best approximation".  We could use any of the $L^p$ norms on the space of continuous functions.  Since the spherical harmonics are infinitely differentiable, a wide variety of other norms are also available.   However we define "best", we can form a basis of the space of cutoff functions, consisting of "pixel $\delta$-functions localized" on the points of an anti-prism distributed over the surface of the sphere.  
  
  We now hypothesize that we can compute the entropy fluctuation by computing the length fluctuation in each of these pixels as independent Gaussian random variables, with covariance computed from the 't Hooft uncertainty relations described above, averaged over the pixel.  Indeed, the area of a pixel is just its angular area, fixed by the number of pixels, times the radius of the screen.  While the number of independent pixels is the same as the number of independent angular momentum components, the form of the fuzzy pixel characteristic functions is not determined by any principle we have been able to invent.    What is clear is that there will be many choices for which
  \begin{equation} \int d^2 n d^2 n^{\prime}  \delta_{LM} (\Omega) \delta_{LM} (\Omega^{\prime}) (J^2 + 1)^{-1} (\Omega, \Omega^{\prime})  = F , \end{equation} is finite as $L_M \rightarrow \infty$.  Here $\delta_{LM}$ is the cutoff $\delta$-function.  
  
  If we now view the fluctuation in area of a large diamond as built up from the fluctuation in area of individual nested diamonds, we can write an {\it area diffusion equation}
  \begin{equation} \frac{dP(A)}{dt} = F A L_P \frac{d^2 P}{d A^2} , \end{equation} for the probability distribution of areas.  The $L_P$ in this equation appears because the time step in our nested sequence of diamonds is the Planck time, and the constant $F$ represents the ambiguity coming from our lack of knowledge of how to regularize the transverse geometry at the Planck scale.   
  
  If we now use the Bekenstein-Hawking-Jacobson-Bousso law for the entropy of the diamond we find that the 't Hooft commutation relations predict area fluctuations that are proportional to the entropy of the diamond, with a proportionality constant that depends on $F$.   Note that in terms of the number of pixels defined above, both the 't Hooft calculation of area fluctuations and the C-S calculation of entropy fluctuations are proportional to the number of pixels, but the C-S calculation contains the ambiguous factor related to the UV cutoff on $1 + 1$ dimensional fermion fields on the stretched horizon.  Recall that this cutoff was also related to UV ambiguities in the Planck scale transverse geometry.  
  
  Both the 't Hooft commutation relations and the C-S ansatz of $1 + 1$ dimensional CFT on the stretched horizon are bold attempts to extract the quantum dynamics of gravity from its classical hydrodynamics, the Einstein field equations.  What we have shown is that, at the level of quadratic fluctuations of entropy, these two approaches are consistent with each other, up to Planck scale ambiguities which we believe are inevitable.  We do not think that these ambiguities can be resolved without experimental input.  The arguments presented in\cite{tbambiguity} suggest that neither perturbative string theory nor AdS/CFT can resolve them.  The C-S approach makes much more far reach assumptions about dynamics on the stretched horizon of causal diamonds, and consequently it makes a more precise prediction for the coefficient in the entropy fluctuations.  We can use this to fix the coefficient $F$ in the length fluctuation calculation.
  
  The connection between length and entropy {\it fluctuations} leads us to a {\it conjecture} about the time correlation function of length fluctuations, namely that it will be given by the length fluctuation in the overlap diamond.   On a very naive level, the argument we gave in the first section for the modular Hamiltonian seems valid also for the length fluctuation, but we do not find this convincing.   
  
  We conclude this section by noting that our approach appears to disagree with that of\cite{VZ3}.  Those authors claim to calculate the modular Hamiltonian of a diamond directly in terms of the 't Hooft operators.   They then argue that, given their formula, both the expectation value and the fluctuation of the modular Hamiltonian are infinite but that the infinities are the same, so that the ratio of the two quantities is $1$.  
  
 \section{Conclusions}
 
 We have shown that the time dependent correlation of modular Hamiltonians in two shifted causal diamonds can be computed in terms of the uncertainty in the modular Hamiltonian of the overlap between the two diamonds.  We then used the 't Hooft commutation relations, together with a regulator scheme inherited from the HST formalism to calculate the length fluctuation in a single diamond, verifying the large value suggested in\cite{VZ1} and consistent with\cite{subsequent}.    Finally, we showed how the length and entropy fluctuation calculations could be related, though here our results were not as precise as one might have hoped.  Nonetheless, taken together they suggest that the length correlation in two shifted diamonds is equal to the length fluctuation in the overlap of the two diamonds and that the logarithmic divergence in the latter is rendered finite in a way that can be calculated from the C-S fluctuation formula for entropy.   This should lead to predictions for the Power Spectral Density in interferometer experiments.

\begin{center}
Acknowledgements 
\end{center} 

  We thank Kathryn Zurek for sharing her insights about fluctuations in causal diamonds.  The work of TB is supported in part by the DOE under grant DE-SC0010008.  The work of WF is supported by the NSF under grant PHY-1914679


\begin{thebibliography}{99}
\bibitem{VZ1} E.~P.~Verlinde and K.~M.~Zurek, ``Observational signatures of quantum gravity in interferometers,''
Phys. Lett. B \textbf{822}, 136663 (2021)
doi:10.1016/j.physletb.2021.136663
[arXiv:1902.08207 [gr-qc]].

\bibitem{subsequent}E.~Verlinde and K.~M.~Zurek,
``Spacetime Fluctuations in AdS/CFT,''
JHEP \textbf{04}, 209 (2020)
doi:10.1007/JHEP04(2020)209
[arXiv:1911.02018 [hep-th]];
Y.~Zhang and K.~M.~Zurek,
``Stochastic description of near-horizon fluctuations in Rindler-AdS,''
Phys. Rev. D \textbf{108}, no.6, 066002 (2023)
doi:10.1103/PhysRevD.108.066002
[arXiv:2304.12349 [hep-th]];
D.~Li, V.~S.~H.~Lee, Y.~Chen and K.~M.~Zurek,
``Interferometer response to geontropic fluctuations,''
Phys. Rev. D \textbf{107}, no.2, 024002 (2023)
doi:10.1103/PhysRevD.107.024002
[arXiv:2209.07543 [gr-qc]];
K.~M.~Zurek,
``On vacuum fluctuations in quantum gravity and interferometer arm fluctuations,''
Phys. Lett. B \textbf{826}, 136910 (2022)
doi:10.1016/j.physletb.2022.136910
[arXiv:2012.05870 [hep-th]];
Y.~Zhang and K.~M.~Zurek,
``Stochastic description of near-horizon fluctuations in Rindler-AdS,''
Phys. Rev. D \textbf{108}, no.6, 066002 (2023)
doi:10.1103/PhysRevD.108.066002
[arXiv:2304.12349 [hep-th]];
T.~He, A.~M.~Raclariu and K.~M.~Zurek,
``From Shockwaves to the Gravitational Memory Effect,''
[arXiv:2305.14411 [hep-th]];
K.~M.~Zurek,
``Snowmass 2021 White Paper: Observational Signatures of Quantum Gravity,''
[arXiv:2205.01799 [gr-qc]].
\bibitem{tHcr} G.~'t Hooft,``Quantum mechanics at the black hole horizon,''
G.~'t Hooft,``More on the black hole S matrix,''
G.~'t Hooft, ``The black hole interpretation of string theory,''
Nucl. Phys. B \textbf{335}, 138-154 (1990)
doi:10.1016/0550-3213(90)90174-C;
G.~'t Hooft,``The Black hole horizon as a quantum surface,''
Phys. Scripta T \textbf{36}, 247-252 (1991)
doi:10.1088/0031-8949/1991/T36/026.
\bibitem{VZ3} E.~Verlinde and K.~M.~Zurek, ``Modular fluctuations from shockwave geometries,''
Phys. Rev. D \textbf{106}, no.10, 106011 (2022)
doi:10.1103/PhysRevD.106.106011
[arXiv:2208.01059 [hep-th]].
\bibitem{Carlip}  S.~Carlip,
``Black hole entropy from horizon conformal field theory,''
Nucl. Phys. B Proc. Suppl. \textbf{88}, 10-16 (2000)
doi:10.1016/S0920-5632(00)00748-9
[arXiv:gr-qc/9912118 [gr-qc]];
S. Carlip, Phys. Rev. D 51, 632 (1995), arXiv:gr-qc/9409052'S. Carlip, Phys. Rev. Lett. 82, 2828 (1999), arXiv:hep-th/9812013, S. Carlip, Class. Quant. Grav. 15, 3609 (1998), arXiv:hep-th/9806026,S. Carlip, AIP Conf. Proc. 1483, 54 (2012), arXiv:1207.1488 [gr-qc].
\bibitem{solo} S. N. Solodukhin, Phys. Lett. B 454, 213 (1999), arXiv:hep-th/9812056, and references therein.

\bibitem{BZ} T.~Banks and K.~M.~Zurek,
``Conformal Description of Near-Horizon Vacuum States,''
[arXiv:2108.04806 [hep-th]].
\bibitem{VZ2}E.~Verlinde and K.~M.~Zurek,``Spacetime Fluctuations in AdS/CFT,''
JHEP \textbf{04}, 209 (2020)
doi:10.1007/JHEP04(2020)209
[arXiv:1911.02018 [hep-th]].
\bibitem{perl} E.~Perlmutter, ``A universal feature of CFT R\'enyi entropy,''
JHEP \textbf{03}, 117 (2014)
doi:10.1007/JHEP03(2014)117
[arXiv:1308.1083 [hep-th]].
\bibitem{page} D.~N.~Page, ``Average entropy of a subsystem,''
Phys. Rev. Lett. \textbf{71}, 1291-1294 (1993)
doi:10.1103/PhysRevLett.71.1291
[arXiv:gr-qc/9305007 [gr-qc]].
\bibitem{hilbertbundles} T.~Banks,``Hilbert Bundles and Holographic Space-time Models,''
[arXiv:2306.07038 [hep-th]].
\bibitem{V293} E.~P.~Verlinde and H.~L.~Verlinde,``High-energy scattering in quantum gravity,''
Class. Quant. Grav. \textbf{10}, S175-S184 (1993)
doi:10.1088/0264-9381/10/S/018
H.~L.~Verlinde and E.~P.~Verlinde,``Scattering at Planckian energies,''
Nucl. Phys. B \textbf{371}, 246-268 (1992)
doi:10.1016/0550-3213(92)90236-5
[arXiv:hep-th/9110017 [hep-th]].
\bibitem{connes} A. Connes, {\it Non-commutative Geometry}, Academic Press, New York 1994, ISBN0-12-185860-X, page 539 {\it et. seq.}, and references therein.  
\bibitem{holocosm} T.~Banks, W.~Fischler and L.~Mannelli,
``Microscopic quantum mechanics of the p = rho universe,''
Phys. Rev. D \textbf{71}, 123514 (2005)
doi:10.1103/PhysRevD.71.123514
[arXiv:hep-th/0408076 [hep-th]];
T.~Banks and W.~Fischler,
Int. J. Mod. Phys. D \textbf{27}, no.14, 1846005 (2018)
doi:10.1142/S0218271818460057
[arXiv:1806.01749 [hep-th]].
\bibitem{tbambiguity} T.~Banks,``On the Impossibility of Precise Verification of Models of Quantum Gravity,''
[arXiv:2309.07203 [gr-qc]].
\end{thebibliography}
\end{document}